\documentclass[11pt]{JHEP3}
\usepackage{axodraw, amsmath,graphicx}
\usepackage{array}

\title{\textbf{Spectral Flow of the Non-Supersymmetric Microstates of
    the D1-D5-KK System}} 
\author{Jassem H. Al-Alawi\thanks{e-mail address: J.H.Al-Alawi@durham.ac.uk}
and Simon F. Ross\thanks{email address: S.F.Ross@durham.ac.uk}  \\
Centre for Particle Theory, Department of Mathematical Sciences, \\
Durham University, Durham DH1 3LE, UK\\
}

\date{\today}

\maketitle

\abstract{We show that a realisation of spectral flow as a coordinate
  transformation for asymptotically four-dimensional solutions can be
  extended to the non-supersymmetric case. We apply this
  transformation to smooth geometries describing microstates of the
  D1-D5-KK monopole system in type IIB supergravity compactified on a
  six-torus, and obtain solutions with an additional momentum
  charge. We study the supersymmetric and near-core limits of this
  construction.}

\begin{document}

\section{Introduction}

Supersymmetric D-brane systems with a large degeneracy of ground
states have been a central element in progress in string theory as a
quantum theory of gravity. One of the theory's great successes is the
discovery that counting such states at weak coupling reproduces the
black hole entropy \cite{SV96,CM96}. The AdS/CFT correspondence
discovered by considering a near-horizon limit of the brane systems
\cite{Mald} then led to a significantly better understanding of the
relation between geometry and D-brane descriptions. Further
understanding of this relation has come from the construction of
smooth horizon-free geometries corresponding to individual states of
the D-brane system, showing that the map between geometry and field
theory can go beyond the thermodynamic regime
\cite{conical,mm,lminfo,lmm}. Though a generic microstate is expected
to admit a description only in the full string theory (as found
recently in \cite{deBoer}), large classes of geometries dual to
supersymmetric microstates of a three-charge brane system in five
dimensions and a four-charge brane system in four dimensions have been
constructed.  Mathur and his collaborators, in a series of papers,
have argued that the information paradox could be resolved if the
black hole geometry is viewed as a coarse grained description,
averaging over geometries describing the individual microstates which
differ in a `fuzz ball' region inside the would-be horizon of the
black hole. See \cite{Mrev,BWrev} for reviews of this work.

To test this proposed description of black holes and to further
improve our understanding of the relation between geometry and
D-branes, it is useful to construct smooth geometries dual to the
non-supersymmetric excited states of the D-brane systems. This allows
us to consider dynamical issues involving transitions between
different states. Very few such geometries have been
constructed. Geometries dual to two- and three-charge brane systems
which are asymptotically flat in five dimensions were constructed in
\cite{JMRT05}, and geometries dual to three-charge brane systems which
are asymptotically flat in four dimensions were constructed in
\cite{GLR07,GRS07}. These geometries are smooth in the duality frame
where the brane charges correspond to a D1-D5-P system compactified on
$T^5$ in the five-dimensional case, and to a D1-D5-Kaluza-Klein
monopole (KKM) system in the four-dimensional case. Although smooth
geometries have been constructed only for very special
non-supersymmetric states, this has already led to interesting
physics; these geometries are unstable \cite{instab}, and this
instability can be precisely reproduced by studying the dynamics of
the corresponding quantum state of the brane system \cite{cm}.

A next step in the construction of smooth non-supersymmetric
geometries would be to obtain smooth solutions which are
asymptotically flat in four dimensions corresponding to a four-charge
D1-D5-P-KKM system. This extension was not undertaken in \cite{GRS07}
because the additional charge seemed to lead to daunting additional
complexities. The aim of this paper is to show that the solution with
D1-D5-P-KKM charges is in fact related to the D1-D5-KKM solution
obtained in \cite{GRS07} by a coordinate transformation, using the
analogue of the transformation studied in \cite{BBW08} in the
supersymmetric case.

This coordinate transformation is related to spectral flow, which has
played an important part in the understanding of these smooth
geometries from the outset. When the first simple examples of smooth
geometries were constructed in \cite{conical}, they were related to
states obtained by spectral flow from the Neveu-Schwarz ground state
in the dual CFT, and this spectral flow was identified with a
coordinate transformation of the near-core AdS$_3$ region in the
spacetime geometry. This coordinate transformation was then exploited
in the construction of further examples of solutions which were
asymptotically flat in five dimensions \cite{sflow}. The near-core
regions of these solutions were related by a coordinate
transformation, but the full asymptotically flat solutions were
physically distinct. The remarkable realisation of \cite{BBW08} is
that once we compactify a further direction on a circle by adding a
Kaluza-Klein monopole charge to obtain solutions which are
asymptotically flat in four dimensions, the spectral flow
transformation which preserves supersymmetry is realised as a
coordinate transformation for the full asymptotically flat solution.

In this paper, we will consider the analogue of the coordinate
transformation used in \cite{BBW08} for the non-supersymmetric
geometries considered in \cite{GRS07}. For non-supersymmetric
geometries, we can consider acting with spectral flow independently on
the left and on the right; these transformations are related to two
independent coordinate transformations in the near-core AdS$_3$
region. However, we find that as in the supersymmetric case, only a
single combination extends to a coordinate transformation of the full
asymptotically flat solution. The coordinate transformation is
labelled by a single integer parameter. This coordinate transformation
adds an additional momentum charge to the solutions, and we show that
it reproduces precisely the expected D1-D5-P-KKM solutions,
corresponding to the three-charge D1-D5-P solutions of \cite{JMRT05}
sitting at the core of a Kaluza-Klein monopole. 

In section \ref{bw}, we review the spectral flow of \cite{BBW08} in the
supersymmetric case. In section \ref{geom}, we give a brief review of
the non-supersymmetric geometries of \cite{GRS07}. Section \ref{sf}
contains the main result of our paper, showing that a spectral flow
coordinate transformation can be used to obtain the expected
D1-D5-P-KKM solution. 

\section{Spectral Flow for Supersymmetric solutions}
\label{bw}

In this section we review the results of \cite{BBW08}, who considered
spectral flow for the supersymmetric solutions which are
asymptotically flat in six dimensions. We will review the aspects of
their analysis which act as an inspiration for the non-supersymmetric
case we consider. Consider therefore a solution of type IIB string
theory compactified on a $T^4$, with D1-brane, D5-brane, momentum and
Kaluza-Klein monopole charges. The ten-dimensional metric is 
\begin{equation}
 ds_{10}^{2}=ds_{6}^{2} + \sqrt{\frac{Z_{1}}{Z_{5}}}ds_{T^{4}},
\end{equation}
where $Z_{1}$ and $Z_{5}$ correspond to the D1 and D5 charges
respectively, and the six-dimensional supergravity metric is a
fibration over a four-dimensional Gibbons-Hawking base,
\begin{equation}
 ds^{2}_{6}=-\frac{2}{H}\left(dv+\beta\right)\left(du+k+\frac{1}{2}F\left(dv+\beta\right)\right)+Hh_{\mu\nu}dx^{\mu}dx^{\nu} 
\end{equation}
where
 \begin{equation}
H=\sqrt{Z_{1}Z_{5}},
 \end{equation}
 and $F$ is a Kaluza-Klein gauge potential. A linear combination of
 $u$ and $v$ parametrises a spatial circle with finite proper size at
 large distance. We choose coordinates such that $v$ is periodic with
 period $2\pi$. We assume that the four-dimensional base
space is a Gibbons-Hawking space, with metric
\begin{equation} \label{ghbase}
 ds_{4}^{2}=h_{\mu\nu}dx^{\mu}dx^{\nu}=V^{-1}\left(d\tau+A\right)^{2}
 +V\gamma^{ij}d\,x^{i}dx^{j},  
\end{equation}
where $i,j=1,2,3$. We assume $\partial_\tau$ is Killing, and $\tau$ is
periodic with period $4\pi$. The one-form gauge fields $k$ and $\beta$
can then be decomposed as
\begin{equation}
 k=\mu\left(d\tau+A\right)+\omega, \quad
 \beta=\nu\left(d\tau+A\right)+\sigma, 
\end{equation}
where $\omega, \sigma$ are one-forms on the three-dimensional
base. This metric solves the IIB supergravity equations of motion with
a two-form RR gauge potential carrying the D1- and D5-brane charges;
see \cite{BBW08} for further details on the equations of motion.

We are interested in solutions where the Gibbons-Hawking metric
\eqref{ghbase} carries a Kaluza-Klein monopole charge, so that the
circle along $\tau$ has finite proper size at large distances, and the
spacetime is asymptotically flat in four dimensions. Since there are
two finite circles at large distances, we can consider
$SL(2,\mathbb{Z})$ coordinate transformations which mix up these two
circles. In particular, \cite{BBW08} consider the coordinate
transformation 
\begin{equation}
 \tau\rightarrow\tau+\gamma v,
\end{equation}
where $\gamma$ is an even integer, so this transformation is
consistent with the periodicities of the coordinates.  From the
six-dimensional point of view, this is simply a coordinate
transformation. It therefore naturally preserves the regularity and
the asymptotic structure of the six-dimensional metric. Since the
transformation changes the identity of the Kaluza-Klein $v$ circle,
from the five-dimensional point of view, this is a highly non-trivial
solution-generating transformation which mixes up the Kaluza-Klein
gauge field with the five-dimensional metric. From the point of view
of a CFT description of the low-energy degrees of freedom on the D1-D5
system, this mixes charges which would be interpreted as R-charges
with the Virasoro generators $L_0, \bar L_0$, so it is naturally
interpreted as a spectral flow automorphism of the CFT.\footnote{
Although the transformation is purely a coordinate transformation in
the six-dimensional description, it will modify the asymptotic moduli
of the solution. Thus, with boundary conditions that fix the
asymptotic metric, this coordinate transformation is a global symmetry
of the theory, rather than a gauge symmetry, and we can think of the
solutions it relates as physically distinct.}

A particular example considered in \cite{BBW08}
is to use this spectral flow to map a two-charge supertube to a
bubbling three-charge geometry. Since the general two-charge supertube
solution has an arbitrary profile for the tube, this can be used to
construct new infinite families of three-charge solutions. Our
interest in the non-supersymmetric case is in the analogue of the
simplest case, when we consider a round supertube and the
corresponding three-charge solution.

\section{The Non-supersymmetric Microstates of the D1-D5-KK
    System}
\label{geom}

We will now briefly review the structure of the smooth
non-supersymmetric solutions carrying D1- D5- and Kaluza-Klein (KK)
monopole charges constructed in \cite{GRS07}, which we want to apply
this transformation to. These solutions are asymptotically flat in
four dimensions, and there is a family of smooth solutions labelled by
a single integer parameter. We will write down the geometry
corresponding to the smooth solutions.

The solutions were constructed by starting from a Kerr-Bolt instanton
and adding the KK monopole charge using an $SL(3,\mathbb{R})$
solution-generating transformation. Then by a sequence of boosts and
dualities, the D1 and D5 charges were added. The resulting geometry is
not smooth in general; restricting to the case where the solution has
no horizons and is everywhere smooth leaves us with a family of
solutions labeled by the D1, D5 and KK monopole charges, the
asymptotic size of the circle the D1 and D5 branes wrap, and a single
integer parameter. The solutions also carry a KK electric charge, but
this is determined by the other charges. The smoothness conditions
also fix the periodic identifications on the coordinates. The metric
for the smooth solutions is
\begin{eqnarray}
 ds^{2}_{10}&=&\left(\tilde H_{1}\tilde H_{5}\right)^{-1/2}\left[ A\left(dy + s_{1}s_{5}\mathcal{B}\right)^{2}-G\left(dt+c_{1}c_{5}\mathcal{A}\right)^{2}\right] \nonumber\\
&&+\left(\tilde H_{1}\tilde
  H_{5}\right)^{1/2}\left[\frac{f^{2}}{AG}\left(dz+\omega^{1}\right)^{2}+\frac{d\rho^{2}}{\Delta}+d\theta^{2}+\frac{\Delta}{f^{2}}\sin^{2}\theta
  d\phi^{2}\right] \nonumber\\
&& +\left(\frac{\tilde H_{1}}{\tilde H_{5}}\right)ds^{2}_{T^{4}} \label{nsmet}
\end{eqnarray}
where
\begin{equation}
 \mathcal{A}=\omega^{0}-\frac{C}{G}\left(dz+\omega^{1}\right),
\end{equation}
\begin{equation}
\mathcal{B}=-V_{0}\left(dz+\omega^{1}\right)+\kappa^{1}_{0},
\end{equation}
\begin{equation}
\tilde H_{1,5}=A+\left(A-G\right)s^{2}_{1,5},
\end{equation}
\begin{equation}
G=A\left(1-H\right)=\frac{Af^{2}-C^{2}}{B}.
\end{equation}
The metric functions are 
\begin{equation}
\Delta=\rho^{2}-\rho_{0}^{2},
\end{equation}
\begin{equation}
  f^{2}=\left(\rho^{2}-\rho_{0}^{2}\right)+\rho^{2}_{0} n^{2} \sin^{2}\theta,
\end{equation}
\begin{equation}
  A= f^{2}+2 p\left[\left(\rho-\rho_{0}\right)+ n^{2}\rho_{0}\left(1+ \cos\theta\right)\right],
\end{equation}
\begin{equation}
  B= f^{2}+2\frac{\rho_{0}\left( p+\rho_{0}\right)\left(
  n^{2}-1\right)}{\left( p-\rho_{0}\left(
  n^{2}-1\right)\right)}\left[\left(\rho-\rho_{0}\right)+
  n^{2}\rho_{0}\left(1- \cos\theta\right)\right], 
\end{equation}
\begin{equation}
  C=\frac{2\rho_{0}\sqrt{\rho_{0}\left(\rho_{0}+ p\right)}} n\left(
  n^{2}-1\right){\left( p-\rho_{0}\left(
  n^{2}-1\right)\right)}\left[\left(\rho-\rho_{0}\right)+\left(\rho_{0}+
  p\right)\left(1- \cos\theta\right)\right], 
\end{equation}
\begin{equation}
 \omega^{0}=\frac{2 J \sin^{2}\theta\left(\rho-\rho_{0}\right)}{ f^{2}} d\phi,
\end{equation}
\begin{equation}
 \omega^{1}=\frac{2}{ f^{2}}\sqrt{ p\left( p-\rho_{0}\left(
 n^{2}-1\right)\right)}\left[\left(\rho^{2}-\rho_{0}^{2}\right)
 \cos\theta-\frac{\rho_{0} p n^{2}}{\left( p-\rho_{0}\left(
 n^{2}-1\right)\right)}\left(\rho-\rho_{0}\right) \sin^{2}\theta-
 n^{2}\rho^{2}_{0} \sin^{2}\theta\right] d\phi, 
\end{equation}
and
\begin{equation}
 V_{0}=-\frac{ n\left( n^{2}-1\right)}{
 A}\sqrt{\frac{\rho_{0}^{3}\left( p+\rho_{0}\right)}{ p\left(
 p-\rho_{0}\left( n^{2}-1\right)\right)^{3}}}\left[ f^{2}+2
 p\left(\rho+ p+\left( p+\rho_{0}\right) \cos\theta\right)\right], 
\end{equation}
\begin{equation}
\kappa^{1}_{0}=\frac{2 n\sqrt{\rho_{0}\left(
    p+\rho_{0}\right)}}{\left( p-\rho_{0}\left(
  n^{2}-1\right)\right)}\frac{ \sin^{2}\theta}{
  f^{2}}\left(\begin{array}{cc} & \frac{\rho_{0}\left(
    n^{2}-1\right)}{\left( p-\rho_{0}\left(
    n^{2}-1\right)\right)}\left( p^{2}+2 p\rho_{0}-\rho_{0}^{2}\left(
  n^{2}-1\right)\right)\left(\rho-\rho_{0}\right) \\ 
&+2\rho_{0}^{2}\left( n^{2}-1\right)\left(\rho_{0}+
  p\right)\end{array}\right), 
\end{equation}
\begin{equation}
 \kappa^{0}_{0}=-\frac{2}{ f^{2}}\frac{\rho_{0}\left(
 p+\rho_{0}\right)\left( n^{2}-1\right)}{\left( p-\rho_{0}\left(
 n^{2}-1\right)\right)}\left[\left(\rho^{2}-\rho_{0}^{2}\right)
 \cos\theta+\frac{}{}\left( p\rho-\rho_{0}^{2}\left(
 n^{2}-1\right)\right) \sin^{2}\theta\right], 
\end{equation}
where 
\begin{equation}
  J^{2}=\frac{\rho^{3}_{0} p\left(\rho_{0}+ p\right) n^{2}\left(
  n^{2}-1\right)^{2}}{\left( p-\rho_{0}\left( n^{2}-1\right)\right)}. 
\end{equation}
The determinant of the metric is 
\begin{equation}
g=-\frac{\tilde H_{1}^{3}}{\tilde H_{5}}\sin^{2}\theta.
\end{equation}
It is convenient to introduce the combinations
\begin{equation}
P^{2}=\frac{p\left(p^{2}+m^{2}\right)}{\left(p+q\right)}, \quad Q^{2}=\frac{q\left(q^{2}+m^{2}\right)}{\left(q+p\right)},
\end{equation}
where
\begin{equation}
q = \frac{ \rho_0 (p+\rho_0) (n^2-1)}{(p-\rho_0(n^2-1))}.
\end{equation}
The charges of the four-dimensional asymptotically flat solution are then
\begin{eqnarray}
 &&\mathcal{M}=\frac{1}{2}\left[p+q\left(1+s^{2}_{1}+s^{2}_{5}\right)\right]\nonumber\\
&&\mathcal{P}=P, \quad \mathcal{Q}=Qc_{1}c_{5}, \nonumber\\
&&\mathcal{J}=Jc_{1}c_{5}, \quad \mathcal{Q}_{i}=qs_{i}c_{i}, \quad i=1,5,
\end{eqnarray}
where $\mathcal{P},\mathcal{Q},\mathcal{Q}_{1}$ and $\mathcal{Q}_{5}$
are the KK monopole, KK electric, D1 and D5 charges.

The metric has coordinate singularities at $\rho=\rho_0$ and
$\theta=0,\pi$. The determinant of the metric at constant $\rho$ and
$t$ is
\begin{equation}
g_{(\rho t)} = \frac{\tilde H_1^2}{\tilde H_5^2 f^2} (\rho-\rho_0)
\sin^2 \theta [ A (\rho+\rho_0) (B c_1^2 c_5^2 - f^2 (c_1^2 s_5^2 +
  s_1^2 c_5^2) + \frac{G f^2}{A} s_1^2 s_5^2) - 2J c_1^2 c_5^5], 
\end{equation}
so at these singularities, a spatial isometry direction is
degenerating. To make these singularities smooth origins, we need to
make appropriate identifications so that the direction which is
degenerating is compact with an appropriate period. This imposes the
identifications 
\begin{equation} \label{ids}
(y,z,\phi) \sim (y, z-4\pi {\mathcal P}, \phi + 2\pi) \sim (y, z+4\pi
  {\mathcal P}, \phi + 2\pi) \sim (y - 2\pi n R_y, z + 4\pi n
  {\mathcal P}, \phi + 2\pi n), 
\end{equation}
where 
\begin{equation}
R_y = 4 q \frac{ \sqrt{q} \sqrt{p+q}}{\sqrt{q^2+m^2}} s_1 s_5.  
\end{equation}
The first two identifications in \eqref{ids} guarantee smoothness at
$\theta=0,\pi$, and the last guarantees smoothness at $\rho=\rho_0$.

To facilitate the comparison to the supersymmetric case, we introduce
``light-cone coordinates'' $u,v$ defined by
\begin{equation}
t=\frac{1}{\sqrt{2}}\left(u+v\right),\quad y=\frac{1}{\sqrt{2}}\left(u-v\right),
\end{equation}
In these coordinates,
\begin{eqnarray}
 ds^{2}_{10}&=&\left(\tilde H_{1}\tilde H_{5}\right)^{- 1/2}\left(\frac{ A- G}{2}\right)\left[\left( du+\beta\right)^{2}+
\left( dv+\omega\right)^{2}\right] \nonumber\\
&& -\left( \tilde H_{1} \tilde H_{5}\right)^{- 1/2}\left( A+ G\right)\left( dv+\omega\right)\left( du+\beta\right)\nonumber\\
&&+\left( \tilde H_{1} \tilde H_{5}\right)^{ 1/2}\left[\frac{ f^{2}}{ A G}\left( dz+\omega^{1}\right)^{2}+\frac{ d\rho^{2}}{\Delta}+ d\theta^{2}+\frac{\Delta}{ f^{2}} \sin^{2}\theta d\phi^{2}\right] \nonumber\\
&& +\left(\frac{ \tilde H_{1}}{ \tilde H_{5}}\right) ds^{2}_{ T^{4}},
  \label{nslc}
\end{eqnarray}
where we define 
\begin{equation}
\zeta_{1}= s_{1} s_{5}\mathcal{ B}=\frac{1}{\sqrt{2}}\left(\beta-\omega\right), \quad \zeta_{2}= c_{1} c_{5}\mathcal{ A}=\frac{1}{\sqrt{2}}\left(\beta+\omega\right),
\end{equation}
and $\beta$ and $\omega$ are given by
\begin{equation}
\beta=\frac{1}{\sqrt{2}}\left(\zeta_{1}+\zeta_{2}\right)=-\eta_{1}\left( dz+\omega^{1}\right)+\eta_{2},
 \end{equation}
\begin{equation}
 \omega=\frac{1}{\sqrt{2}}\left(\zeta_{2}-\zeta_{1}\right)=\eta_{3}\left( dz+\omega^{1}\right)+\eta_{4},
\end{equation}
where
\begin{equation}
 \eta_{1}=\frac{1}{\sqrt{2}}\left( s_{1} s_{5} V_{0}+ c_{1} c_{5}\frac{ C}{ G}\right),
\end{equation}
\begin{equation}
 \eta_{2}=\frac{1}{\sqrt{2}}\left( c_{1} c_{5}\omega^{0}+ s_{1} s_{5}\kappa^{1}_{0}\right),
\end{equation}
\begin{equation}
\eta_{3}=\frac{1}{\sqrt{2}}\left( s_{1} s_{5} V_{0}- c_{1} c_{5}\frac{ C}{ G}\right),
\end{equation}
\begin{equation}
 \eta_{4}=\frac{1}{\sqrt{2}}\left( c_{1} c_{5}\omega^{0}- s_{1} s_{5}\kappa^{1}_{0}\right).
\end{equation}

\subsection{BPS case}

The solution is supersymmetric for $n=1$, where $m=0$ and we must take
the $\delta_i \to \infty$ to hold the charges ${\mathcal Q}_i$
fixed. This case therefore requires a slightly separate discussion. 
When $n=1$, $C=0$ and $B= f^2 = \rho^2 - \rho_0^2 \cos^2 \theta$ , so
$G=A$, but  
\begin{equation}
\left( 1- \frac{G}{A} \right) \sinh^2 \delta_i = \frac{2 {\mathcal
    Q}_i}{\rho+\rho_0 \cos \theta},
\end{equation}
so $\tilde H_i = A \left( 1 + \frac{2 {\mathcal Q}_i}{\rho+\rho_0
  \cos \theta} \right)$. Similarly, the one-forms $\omega$ and $\beta$ in
  \eqref{nslc} will have finite limits. We also have 
\begin{equation}
\frac{A}{f^2} = 1 + \frac{2p}{\rho - \rho_0 \cos \theta}. 
\end{equation}
It is then useful to introduce the new coordinates
\begin{equation}
  \tilde r=\rho- \rho_0 \cos\theta, \quad  \cos\tilde\theta=\frac{\rho
  \cos\theta- \rho_0}{\rho- \rho_0 \cos\theta} 
\end{equation}
and define the parameters
\begin{equation}
c=2 b, \quad  Q_{K}=2\mathcal{ P}, \quad  Q_{Ke}=2\mathcal{ Q},
\quad  Q_{i}=2\mathcal{Q}_{i}, \quad  i=1,5. 
\end{equation}
We then have
\begin{equation}
V= \frac{A}{f^2} = 1 + \frac{Q_K}{\tilde r}, \quad Z_i = \frac{\tilde
  H_i}{A} = 1 + \frac{Q_i}{\tilde r_c}, \quad i=1,5
\end{equation}
where 
\begin{equation}
\tilde r_{ c}=\sqrt{ \tilde r^{2}_{ c}+ c^{2}+2 c \tilde r \cos\tilde\theta}.
\end{equation}
The metric \eqref{nslc} in the supersymmetric case $n=1$ then takes
the form
\begin{eqnarray}
ds^{2}_{10}&=&-\frac{2}{ H}\left( d v+\omega\right)\left( d u+\beta\right)+ H V^{-1}\left( d z+\omega^{1}\right)\nonumber\\
&&+ H V\left( d \tilde r^{2}+ \tilde r^{2} d\tilde\theta^{2}+ \tilde
  r^{2} \sin^{2}\tilde\theta^{2} d\phi^{2}\right)+\left(\frac{ Z_{1}}{
    Z_{5}}\right) ds^{2}_{ T^{4}}, \label{bpsm}
\end{eqnarray}
where $ H=\sqrt{ Z_{1} Z_{5}}$, reproducing the form used for example
in \cite{BBW08}. Note however that the coordinates here are not
exactly the same as in \cite{BBW08}; in particular, by \eqref{ids}, in
\eqref{bpsm} the $z$ coordinate has period $4\pi Q_K$, and the $v$
coordinate has period $2\pi n R_y$. 

\subsection{\bf{Near-Core Limit}}
\label{ncore}

The solution reviewed above is supposed to be interpreted as the
familiar smooth D1-D5 solution of \cite{JMRT05} which is asymptotically
flat in five dimensions, sitting at the core of a Kaluza-Klein
monopole which converts it into a solution which is asymptotically
flat in four dimensions. In \cite{GRS07}, this was argued by showing
that the metric \eqref{nsmet} has a near-core limit where it reduces to an
AdS$_3 \times S^3$ geometry of the same form as is obtained in the
near-core limit of the five-dimensional solution of \cite{JMRT05}. We
will now briefly review this near-core limit. 

The appropriate limit is to take $\rho_0 \to 0$ holding $p$ and the
D1, D5 brane charges ${\mathcal Q}_i$ fixed. This limit will scale
$\mathcal{ Q}$ and $\mathcal{ J}$ to zero, so it is distinct from the
supersymmetric limit. As we take this limit, we scale the coordinates
so as to zoom in on a core region in the geometry, scaling $\rho$ like
$\rho_0$, and the identification on the $y$ coordinate scales like
$1/\sqrt{\rho_0}$. We therefore define new coordinates by 
\begin{equation}
 \rho=\rho_{0} r,  \quad y=\frac{\chi}{4\sqrt{ p\rho_{0}}}, \quad t=\frac{\tau}{4\sqrt{ p\rho_{0}}}, \quad 
z= p\psi, \label{coret1}
\end{equation}
and take the limit keeping $r,\chi,\tau$ finite. In this limit the
metric \eqref{nsmet} becomes
\begin{eqnarray}
&& d s_{10}^{2}\approx \nonumber\\
&&\frac{1}{4\ell^{2}}\left[ a\left[ d\chi+\frac{\ell^{2} n}{2}\left(\frac{\left(1+ \cos\theta\right)}{ a}\left( d\psi+\bar\omega^{1}\right)+\bar\kappa^{1}\right)\right]^{2}- g\left[ d\tau+\frac{\ell^{2} n}{2}\left(\bar\omega^{0}-\frac{\left(1- \cos\theta\right)}{ g}\left( d\psi+\bar\omega^{1}\right)\right)\right]^{2}\right]\nonumber \\
&&+\frac{\ell^{2}}{4}\left[\frac{ d r^{2}}{ r^{2}-1}+ d\theta^{2}+\frac{ r^{2}-1+ n^{2} \sin^{2}\theta}{ a g}\left( d\psi+\bar\omega^{1}\right)^{2}+\frac{\left( r^{2}-1\right) \sin^{2}\theta}{\left( r^{2}-1+ n^{2} \sin^{2}\theta\right)} d\phi^{2}\right]\nonumber \\
&&+\sqrt{\frac{\mathcal{ Q}_{1}}{\mathcal{ Q}_{5}}} d s^{2}_{
    T^{4}}. \label{nsnh} 
\end{eqnarray}
where we have set 
\begin{equation}
 \ell^{2}=4\sqrt{ \tilde H_{1} \tilde H_{5}}=16 p\sqrt{\mathcal Q_{1}\mathcal Q_{5}},
\end{equation}
and
\begin{equation}
  a=2\left( r-1+ n^{2}\left(1+ \cos\theta\right)\right), \quad
  g=2\left( r+1- n^{2}\left(1- \cos\theta\right)\right), 
\end{equation}
\begin{equation}
 \bar\omega^{0}=\frac{\left( r-1\right) \sin^{2}\theta}{ r^{2}-1+
 n^{2} \sin^{2}\theta} d\phi, 
\end{equation}
\begin{equation}
 \bar\omega^{1}=2\frac{\left( r^{2}-1\right) \cos\theta- n^{2} r^{2}
 \sin^{2}\theta}{ r^{2}-1+ n^{2} \sin^{2}\theta} d\phi, 
\end{equation}
\begin{equation}
 \bar\kappa=\frac{\left( r+1\right) \sin^{2}\theta}{ r^{2}-1 +n^{2}
 \sin^{2}\theta} d\phi. 
\end{equation}

This metric has an AdS$_{3}\times S^{3}$ geometry at least locally,
and this can be manifested by introducing new coordinates,
\begin{equation}
  r=1+2 R^{2}, \quad \chi = \ell^2 \varphi, \quad
  \theta=2\bar\theta. \label{coret2}  
\end{equation}
\begin{equation}
\bar{\psi} = \frac{1}{4}(2\phi + \psi), \quad \bar{\phi} =
\frac{1}{4}(2 \phi - \psi). \label{coret3}
\end{equation}
The metric \eqref{nsnh} is then
\begin{eqnarray}
ds^2 &=& - \frac{R^2+1}{\ell^2} d\tau^2 + \frac{\ell^2 dR^2}{R^2+1} +
\ell^2 R^2 d\varphi^2 \\ && + \ell^2 (d \bar{\theta}^2 + \cos^2
\bar{\theta} (d\bar{\psi} + n d\varphi)^2 + \sin^2 \bar{\theta} (d\bar
\phi - \frac{n}{\ell^2} d\tau)^2 ) + \sqrt{\frac{{\mathcal
      Q}_1}{{\mathcal Q}_5}} ds^2_{T^4}. \nonumber 
\end{eqnarray}
In the near-core limit, 
\begin{equation} \label{coreRy}
n R_y = 4 \sqrt{\frac{p}{\rho_0}} \sqrt{{\mathcal Q}_1 {\mathcal Q}_5} ,
\end{equation}
 so the identifications \eqref{ids}
become in these coordinates simply $\bar{\psi} \sim \bar{\psi}+ 2\pi$,
$\bar{\phi} \sim \bar{\phi} + 2\pi$, and $(\varphi, \bar{\psi}) \sim
(\varphi - 2\pi, \bar{\psi}+2\pi n)$. If these are the fundamental
identifications, the spacetime is then globally AdS$_3 \times S^3$.

\section{Spectral Flow}
\label{sf}

Let us now consider the construction of new solutions by acting on the
non-supersymmetric geometry \eqref{nsmet} with a spectral flow
coordinate transformation. As in \cite{BBW08}, we consider the
coordinate transformation 
\begin{equation} \label{sflow}
z \rightarrow z+\gamma v,
\end{equation}
where $\gamma$ is a parameter, and we define the spectral flow using
the coordinates of \eqref{nslc}. It is then clear that the spectral
flow we consider here will coincide with the one studied in
\cite{BBW08} in the supersymmetric case $n=1$, where it corresponds to
the simple example we mentioned in section \ref{bw}, relating the round
two-charge supertube in the Kaluza-Klein monopole background to a
three-charge bubbling solution in the same monopole background.

In general, acting on the non-supersymmetric
metric \eqref{nslc} with the spectral flow \eqref{sflow}, we will
obtain a new solution
\begin{eqnarray}
&&  ds^{2}_{6}=-\frac{2}{ \tilde H}\left( dv+\tilde\omega\right)\left[ du+\tilde\beta+\frac{ \tilde F}{2}\left( dv+\tilde\omega\right)\right]+ \tilde H \tilde V^{-1}\left( dz+\tilde\omega^{1}\right)\nonumber \\
&&+ \tilde H \tilde V\left[( d \tilde r^{2}+ \tilde r^{2} d\tilde\theta^{2}+ \tilde r^{2} sin^{2}\tilde\theta^{2} d\phi^{2}\right],
\end{eqnarray}
where
\begin{equation}
\tilde\omega=\left(1+\gamma\eta_{3}\right)^{-1}\omega, \quad  \tilde H=\left(1+\gamma\eta_{3}\right)^{-1} H, \quad  \tilde V=\left(1+\gamma\eta_{3}\right) V,
\end{equation}
\begin{equation}
\tilde\omega^{1}=\omega^{1}-\gamma\eta_{4}, \quad  \tilde F=-2\gamma\eta_{1}-\frac{\gamma^{2} H^{2} V^{-1}}{\left(1+\gamma\eta_{3}\right)},
\end{equation}
\begin{equation}
\tilde\beta=\beta+\frac{\gamma\eta_{1}}{\left(1+\gamma\eta_{3}\right)} \omega
+\frac{\gamma^{2} H^{2} V^{-1}}{\left(1+\gamma\eta_{3}\right)^{2}} \omega
-\frac{\gamma H^{2} V^{-1}}{\left(1+\gamma\eta_{3}\right)}\left( dz+\omega^{1}\right).
\end{equation}
As in the supersymmetric case, this new solution has an additional
charge given by the Kaluza-Klein gauge potential $\tilde F$,
corresponding to a momentum (Kaluza-Klein electric) charge along the
circle that the D1 and D5 branes wrap. Note that this is not the same
as the Kaluza-Klein electric charge which is already present in the
solution \eqref{nslc}, which is associated with the Kaluza-Klein gauge
field coming from the $z$ circle. We therefore refer to the resulting
solution as a four-charge D1-D5-P-KKM solution. 

In general, this solution will not have the same asymptotics as the
solution that we started with. As in the supersymmetric case, we need
to quantise $\gamma$ to ensure that the spectral flow transformation
preserves the identifications \eqref{ids}. Starting from the
identifications \eqref{ids} in the original coordinates and applying
the transformation $z \to z + \gamma v$, the identifications in the
new coordinates are
\begin{equation} \label{sflowid}
(y,z,\phi) \sim (y, z-4\pi {\mathcal P}, \phi + 2\pi) \sim (y, z+4\pi
  {\mathcal P}, \phi + 2\pi) \sim (y - 2\pi n R_y, z + 4\pi n
  {\mathcal P} - 2\pi n R_y \gamma, \phi + 2\pi n). 
\end{equation}
we want these to be the same as the identifications \eqref{ids} for
the new coordinates. This will be true if $z \sim z - 2\pi n R_y
\gamma$. Since \eqref{sflowid} implies $z \sim z + 8\pi {\mathcal P}$,
this requires 
\begin{equation} \label{gquant}
 \gamma=\frac{4 m\mathcal{ P}}{ n R_{ y}}
\end{equation}
for some integer $m$. This is the analogue for our case of the
restriction of $\gamma$ to even integers in the supersymmetric case, and
taking into account the difference in our normalization of $z, v$, it
will reduce to that identification in the supersymmetric case. 

Spectral flow then gives us a solution labelled by the D1 and D5 brane
and Kaluza-Klein monopole charges, the size of the $y$ circle $R_y$,
and two integer parameters $m,n$.\footnote{We can think equivalently think of
these solutions as labelled by the D1, D5, P charges, the Kaluza-Klein
monopole charge and the two integer parameters $m,n$.} This is exactly
the number of solutions we would expect if we were to put the
three-charge D1-D5-P smooth solutions of \cite{JMRT05} at the core of a
Kaluza-Klein monopole. We will relate the solution here to the
solution of \cite{JMRT05} by considering the near-core limit, as was done
for the solution without the momentum charge in \cite{GRS07}.

Since the near-core limit has already been worked out in the original
coordinates before we perform the spectral flow in \cite{GRS07}, as
reviewed in section \ref{ncore}, all we need to do is to consider the
action of the spectral flow transformation in this near-core
region. Using the coordinate transformations
(\ref{coret1},\ref{coret2},\ref{coret3}), the spectral flow transformation
\eqref{sflow} becomes, in this near-core region,
\begin{equation}
\bar \psi \to \bar \psi +  \frac{\gamma}{16 p \sqrt{p \rho_0}}\left(\tau-\ell^2\varphi\right)
  , \quad \bar \phi \to \bar \phi -\frac{\gamma}{16 p \sqrt{p \rho_0}}\left(\tau-\ell^2\varphi\right).
\end{equation}
In the near-core region, $R_y$ is given by \eqref{coreRy}, so the
quantization condition  \eqref{gquant} becomes
\begin{equation}
\gamma=m\sqrt{p \rho_0}\frac{1}{\sqrt{{\mathcal Q}_1 {\mathcal Q}_5} },
\end{equation}
and $\ell^2 = 16 p \sqrt{{\mathcal Q}_1 {\mathcal Q}_5} $, so the spectral flow
transformation in the near-core region is 
\begin{equation}
\bar \psi \to \bar \psi + \frac{m}{\ell^2}\left(\tau-\ell^2\varphi\right)
  , \quad \bar \phi \to \bar \phi -\frac{m}{\ell^2}\left(\tau-\ell^2\varphi\right)
  .
\end{equation}
This shows that the near-core limit of the spectral flow coordinate
transformation \eqref{sflow} agrees with the usual notion of spectral
flow for AdS$_3 \times S^3$ spacetimes, and the near-core metric of
the new solution obtained by acting with this spectral flow
transformation will be
\begin{eqnarray}
 &&\,d\,s^{2}\approx-\frac{\left(1+\,R^{2}\right)}{\ell^{2}}\,d\tau^{2}+\ell^{2}\,R^{2}\,d\varphi^{2} +\frac{\ell^{2}\,d\,R^{2}}{\,R^{2}+1}+\ell^{2}\,d\bar\theta^{2} \nonumber\\
&&+\ell^{2}\,cos^{2}\bar\theta\left(\,d\bar\psi+\frac{\,m}{\ell^{2}}\,d\tau-\left(\,m-\,n\right)\,d\varphi\right)^{2}+\ell^{2}\,sin^{2}\bar\theta\left(\,d\bar\phi-\frac{1}{\ell^{2}}\left(\,m+\,n\right)\,d\tau+\,m\,d\varphi\right)^{2} \nonumber\\
&&+\sqrt{\frac{\mathcal{\,{Q}}_{1}}{\mathcal{\,Q}_{5}}}\,ds^{2}_{\,T^{4}}.
\end{eqnarray}
This agrees with the near-core region of the three-charge solution in
\cite{JMRT05}, up to a relabelling of the integer parameters and a
trivial shift of $\bar{\psi}, \bar{\phi}$ by terms proportional to
$\tau$. Thus, these four-charge D1-D5-P-KKM solutions obtained by
\eqref{sflow} can indeed be identified with the three-charge solution
of \cite{JMRT05} sitting at the core of a Kaluza-Klein monopole.

Thus, we have shown that the spectral flow coordinate transformation
\eqref{sflow} can be used to construct the remaining simple example of
a non-supersymmetric solution, the D1-D5-P-KKM solution. We could
also consider orbifolds of the solution as in \cite{JMRT05}. Finding more
general non-supersymmetric smooth solutions remains an important open
problem, which will require radically new techniques.

\end{document}